\documentclass[prl,twocolumn]{revtex4}% Physical Review B

\usepackage{graphicx}% Include figure files
\usepackage{dcolumn}% Align table columns on decimal point
\usepackage{bm}% bold math
\def\ä{\"{a}}
\def\ü{\"{u}}
\def\ö{\"{o}}
\def\Ä{\"{A}}
\def\Ü{\"{U}}
\def\Ö{\"{O}}
\begin{document}

\preprint{APS/123-QED}

\title{Local spectroscopy and atomic imaging of tunneling current, forces and dissipation on graphite}% Force line breaks with \\

\author{S. Hembacher}
\author{F. J. Giessibl}
\email{fjg@physik.uni-augsburg.de}
\author{J. Mannhart}

%\altaffiliation[Also at ]{}%Lines break automatically or can be forced with \\
%\author{}%

\affiliation{%
Universit\"at Augsburg, Institute of Physics, Electronic
Correlations and Magnetism, Experimentalphysik VI,
Universit\"atsstrasse 1, D-86135 Augsburg, Germany.
}%
\author{C. F. Quate}
\affiliation{%
Stanford University, Ginzton Lab, Stanford CA 94305-4085, USA.
}%

\date{accepted at PRL}% It is always \today, today,

\begin{abstract}
Theory predicts that the currents in scanning tunneling microscopy
(STM) and the attractive forces measured in atomic force
microscopy (AFM) are directly related. Atomic images obtained in
an attractive AFM mode should therefore be redundant because they
should be \emph{similar} to STM. Here, we show that while the
distance dependence of current and force is similar for graphite,
constant-height AFM- and STM images differ substantially depending
on distance and bias voltage. We perform spectroscopy of the
tunneling current, the frequency shift and the damping signal at
high-symmetry lattice sites of the graphite (0001) surface. The
dissipation signal is about twice as sensitive to distance as the
frequency shift, explained by the Prandtl-Tomlinson model of
atomic friction.

%Valid PACS numbers may be entered using the \verb+\pacs{#1}+
%command.
\end{abstract}

\pacs{68.37.Ef,68.47.Fg,68.37.Ps}% PACS, the Physics and Astronomy
\maketitle

The capability of scanning tunneling microscopy (STM)
\cite{Binnig:1982} and atomic force microscopy (AFM)
\cite{Binnig:1986b} to resolve single atoms in real space makes
them powerful tools for surface science and nanoscience. When
operating AFM in the repulsive mode, protrusions in the images
simply relate to the atoms because of Pauli's exclusion law. In
contrast, the interpretation of STM images is more complicated.
The Tersoff-Hamann approximation \cite{Tersoff:1985}, valid for
tips in an s-state, interprets STM images as a map of the charge
density of the sample at the Fermi energy. Depending on the state
of the tip, atoms can either be recorded as protrusions or holes,
and tip changes can reverse the atomic contrast
\cite{Barth:1990,Chen:1991}. Theoretical predictions regarding the
relation of forces and tunneling currents $I$ state that tunneling
currents and attractive forces are directly related, thus AFM
would not provide any new physical insights over STM. Chen
\cite{Chen:1991} has found that the square of the attractive
energy between tip and sample should be proportional to $I$ with
experimental evidence in \cite{Loppacher:2000b}. Hofer and Fisher
\cite{Hofer:2003} suggested that the interaction energy and $I$
should be directly proportional, experimentally found in
\cite{Schirmeisen:2000,Rubio-Bollinger:2004}. In this Letter, we
investigate the experimental relationships between tunneling
currents and conservative as well as dissipative forces for
graphite probed with a W tip by performing local spectroscopy on
\emph{specific} lattice sites. While force spectroscopy on
specific lattice sites \cite{Lantz:2001} and combined force and
tunneling spectroscopy on unspecific sample positions
\cite{Schirmeisen:2000,Rubio-Bollinger:2004} have been performed
before, the measurements reported here encompass site-specific
spectra of currents \emph{and} forces, supplemented by
simultaneous constant-height measurements of currents and forces
that allow a precise assessment of the validity of the theories
regarding currents and forces in scanning probe experiments.

In graphite (see Fig. 1), the electrons at $E_F$ are concentrated
at the $\beta$ sites, and only these atoms are \lq seen\rq{} by
STM at low-bias voltages.
\begin{figure}
\includegraphics[width=7cm]{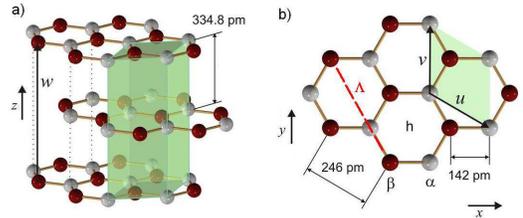}
\caption{(color) Crystal structure of graphite. The unit cell
(green) consists of two layers with inequivalent basis atoms
$\alpha$ (white) and $\beta$ (red). The $\alpha$-atoms have direct
neighbors in the adjacent atomic layers as indicated by the dotted
lines, the $\beta$-atoms are above a hollow site (h). (a)
Perspective view, showing three layers formed by hexagonal rings.
(b) Top view with surface unit vectors $u$ and $v$. The line
$\Lambda=-v+u$ connects the $\alpha$, $\beta$ and h lattice
points, spaced by 142\,pm. }\label{fig1}
\end{figure}
The state-of-the-art method for atomic resolution force microscopy
is frequency modulation AFM (FMAFM) \cite{Albrecht:1991}, where
the frequency shift $\Delta f$ of an oscillating cantilever with
stiffness $k$, eigenfrequency $f_0$ and oscillation amplitude $A$
is used as the imaging signal \cite{Garcia:2002,Giessibl:2003}.
The bonding energy between two adjacent graphite layers at
distance $\sigma$ can be approximated by a Morse potential
\begin{equation}
V_{M}=E_{bond}[-2\exp^{-\kappa(z-\sigma)}+\exp^{-2\kappa(z-\sigma)}]
\end{equation}
with $E_{bond}\approx -23$\,meV and $\kappa\approx 8$\,nm$^{-1}$
per atom pair \cite{Schabel:1992}. The \lq normalized frequency
shift\rq{} $\gamma(x,y,z)=kA^{3/2}\Delta f(x,y,z)/f_0$ connects
the physical observable $\Delta f$ and the underlying forces
$F_{ts}$ with range $\lambda$, where $\gamma\approx 0.4 F_{ts}
\lambda^{0.5}$ (see Eqs. 35-41 in \cite{Giessibl:2003}). For
covalent bonds, the typical bonding strength is on the order of
-1\,nN with $\lambda \approx 1$\,\AA, resulting in $\gamma \approx
-4$\,fN$\sqrt{\textrm{m}}$, where a negative sign indicates
attractive interaction. For graphite, the interlayer bonds are
much weaker and the potential of Eq. 1 results in $\gamma_{min}=
-0.1$\,fN$\sqrt{\textrm{m}}$. The interaction of a tip atom with a
graphite surface may be stronger than the interlayer bonds but
should still result in $-\gamma_{min}<1$\,fN$\sqrt{\textrm{m}}$,
posing a challenge for AFM imaging. True atomic resolution on
graphite by AFM has so far only been obtained at low temperatures,
first by Allers et al. \cite{Allers:1999a} using large-amplitude
FM-AFM with $\Delta f=-63$\,Hz, $f_0=160$\,kHz, $k=35$\,N/m and
$A=8.8$\,nm, thus $\gamma=-11.4$\,fN$\sqrt{\textrm{m}}$. In
spectroscopic measurements by the same group, a minimum of
$\gamma\approx -60$\, fN$\sqrt{\textrm{m}}$ has been observed
(Fig. 1b in \cite{Hoelscher:2000}). Because this value is more
than two orders of magnitude greater than the estimate above, it
is expected that long-range forces have caused a large
contribution in that experiment. Here, we use FMAFM with sub-nm
amplitudes which greatly reduces the influence of long-range
forces \cite{Giessibl:2003} and enables simultaneous STM operation
\cite{Hembacher:2003}.

We use a low-temperature STM/AFM operating at 4.9\,K in ultrahigh
vacuum \cite{Hembacher:2003b}. The microscope uses a qPlus sensor
\cite{Giessibl:1998} for simultaneous STM/AFM operation
($k=1800$\,N/m, $f_0=11851.75$\,Hz, quality factor $Q=20000$). All
data (spectroscopy and images) are recorded at $A=0.25$\,nm. The
tip is prepared by dc etching (3\,V) of a polycrystalline tungsten
wire. The frequency shift is measured with a commercial
phase-locked-loop detector (EasyPLL by Nanosurf AG, Liestal,
Switzerland). The instrument is thermally well connected to a
liquid He bath cryostat at 4.2\,K, leading to a drift rate of
$\approx 20$\,pm/hour. Non-conservative tip-sample interactions
lead to damping, and the energy $\Delta E_{ts}$ that has to be
provided for each oscillation cycle to keep $A$ constant is
recorded simultaneously with $I$ and $\Delta f$.

\begin{figure}
\includegraphics[width=4cm]{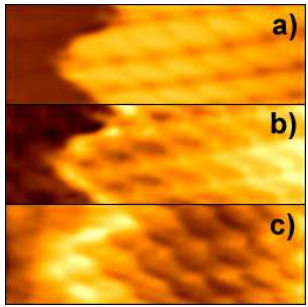}
\caption{ (color) Simultaneous records of a) tunneling current, b)
frequency shift and c) damping. Image size 3\,nm $\times$ 1\,nm,
tip bias 100\,mV, scanning speed 40\,nm/s.} \label{fig2}
\end{figure}

To check if the three data channels $I$, $\Delta f$ and $\Delta
E_{ts}$ are produced by the same tip atom, we have scanned an area
that contains a step edge (Fig. \ref{fig2}).
The step edge appears at the same position in all three data
channels, thus the signals are produced by the same tip atom.

\begin{figure}
\includegraphics[width=6cm]{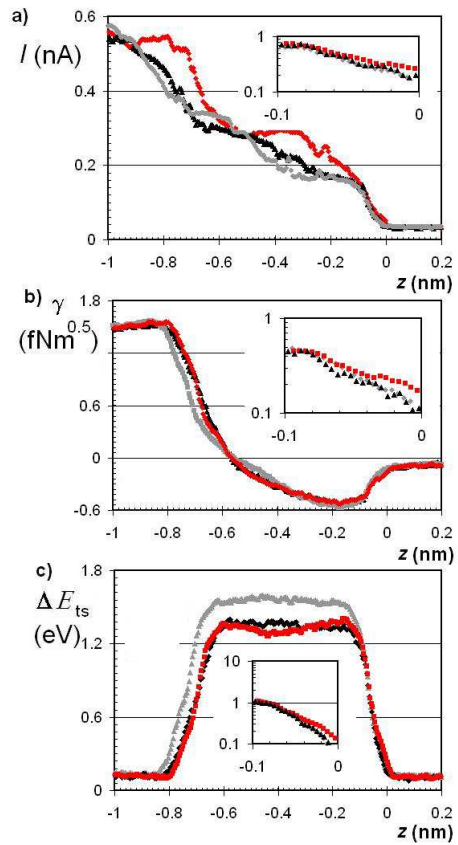}
\caption{ (color) Experimental spectra of $I$, $\gamma$ and
$\Delta E_{ts}$ as a function of distance for a tip bias of
160\,mV. The insets are logarithmic plots of $I$, $-\gamma$ and
$\Delta E_{ts}$. The grey curves are taken at position \lq1\rq{}
in Fig. 4a), the black curves at position \lq2\rq{} and the red
curves at position \lq3\rq. The insets are views of $I(z)$,
$\gamma(z)$ and $\Delta E_{ts}(z)$ for $-0.1$\,nm$<z<0$ on
logarithmic scales.} \label{fig3}
\end{figure}

The physics of the interaction is best explored by performing $I$,
$\gamma$ and $\Delta E_{ts}$-spectroscopy at the high-symmetry
lattice sites. Figure \ref{fig3} shows $I(z)$, $\gamma(z)$ and
$\Delta E_{ts}(z)$ taken at the $\alpha$-, $\beta$- and h-lattice
sites. All three signals initially increase roughly exponentially,
as shown in the log-scale insets. The correspondence between
experimental images and the lattice sites is established by the
analysis below.
The current increases exponentially in the distance regime from
$z=0$ to $z=-100$\,pm, followed by a step-like increase for
smaller distances. We assume that the tip is essentially in
contact to the upper graphite layer for distances smaller than
-100\,pm. Given that the electrical conductivity of graphite is
small at low temperatures and small in $z$-direction, we conclude
that the step-like increase in current for distances smaller than
-100\,pm is caused by an increasing number of graphite layers
becoming available for charge transport. The current spectra in
Fig. 3a) are maximal for most distances on site \lq3\rq. Because
site \lq3\rq{} is a current maximum, we identify it as a $\beta$
position.

The normalized frequency $\gamma$ shift decreases down to a
distance of $\approx -170$\,pm, followed by a slight increase down
to $\approx -550$\,pm and a sharper increase for even smaller
distances. For $z<-800$\,pm, $\gamma$ remains constant because the
cantilever remains in contact to the sample during the entire
oscillation cycle. In the $z$-range from 0 to -100\,pm where the
short-range chemical bonding forces start to emerge, the magnitude
of $\gamma$ increases exponentially (see inset). It is believed,
that the short-range forces between AFM tips and graphite
originate from van-der-Waals forces \cite{Hoelscher:2000} with
their typical $1/z^7$-distance dependence. The experimental data
shows that when the interatomic distance approaches the atomic
diameters, an exponential force dependence prevails.

The damping signal initially also increases exponentially for
$z<0$, reaches a plateau for $z<-170$\,pm, decays to
zero from $z<-600$\,pm and remains zero for $z<-800$\,pm because the
cantilever remains in contact for the whole oscillation cycle. This points to a
damping mechanism as described by Prandtl \cite{Prandtl:1928} and
Tomlinson \cite{Tomlinson:1929}, where the energy loss is caused
by a plucking action of the atoms on each other. The energy loss
per cycle is simply related to the maximal attractive force
$F_{ts\,min}$ and the stiffness of the sample $k_{sample}$ with
$\Delta E_{ts}=F_{ts\,min}^2/(2k_{sample})$ \cite{Giessibl:2001d}.

The insets are logarithmic plots of $I(z)$, $\gamma(z)$ and
$\Delta E_{ts}(z)$ for $-0.1$\,nm$<z<0$, showing an almost
exponential distance dependence in that range. For the tunneling
current, we find a decay constant $\kappa_I=13$\,nm$^{-1}$ at the
$\beta$-site and $\kappa_I=15$\,nm$^{-1}$ at sites \lq 1\rq and
\lq 2\rq, leading to an apparent barrier height of $\approx
2$\,eV. The decay constant of $\gamma$ and thus the interaction
potential (Eq. 39 in \cite{Giessibl:2003}) is
$\kappa_{\gamma}=12$\,nm$^{-1}$ at the $\beta$-site and
$\kappa_{\gamma}=16$\,nm$^{-1}$ at sites \lq 1\rq and \lq 2\rq.
The decay constants for $\kappa_I$ and $\kappa_{\gamma}$ are equal
within the measurement accuracy, thus the theory by Hofer and
Fisher \cite{Hofer:2003} appears to hold for the interaction of W
with graphite. The damping signal decays with $\kappa_{\Delta
E_{ts}}=20$\,nm$^{-1}$  at the $\beta$-site and $\kappa_{\Delta
E_{ts}}=30$\,nm$^{-1}$ at sites \lq 1\rq and \lq 2\rq{} as
expected from an energy loss proportional to the square of the
attractive force.

The normalized frequency spectra shown in Fig. 3b) are rather
similar, expect for the grey curve recorded at site \lq1\rq. In
the distance regime from $-600$\,pm to $-800$\,pm, the grey curve
is shifted by $\approx -40$\,pm. If the C sample atoms and the W
tip atom are assumed to be hard spheres with a diameter of 142\,pm
and 273\,pm respectively, the W tip atom could protrude 56\,pm
deeper on top of the hollow sites than at $\alpha$ or $\beta$
sites. We therefore conclude that position \lq1\rq{} (grey curves)
corresponds to a hollow site (h in Fig. 1(b)). Interestingly, Fig.
3c) shows that the dissipation is significantly larger on the
hollow site than on top of $\alpha$ or $\beta$ sites. The three
data channels were acquired simultaneously and the range from
$z=-1.7$\,nm to $0.6$\,nm and back to $-1.7$\,nm was ramped within
60\,s. The sequence of the three sites was scanned three times, so
a total of 6 spectra was collected for each $\alpha$-, $\beta$-
and h-site. While the drift rate of our instrument is very low,
piezo creep caused $z$-offsets of consecutive scans. These offsets
were calibrated by comparison with constant-height scans which
provide precise cross references for $I$, $\gamma$ and $\Delta
E_{ts}$ at the $\alpha$-, $\beta$- and h-sites for a given
$z$-value.
\begin{figure}
\includegraphics[width=8cm]{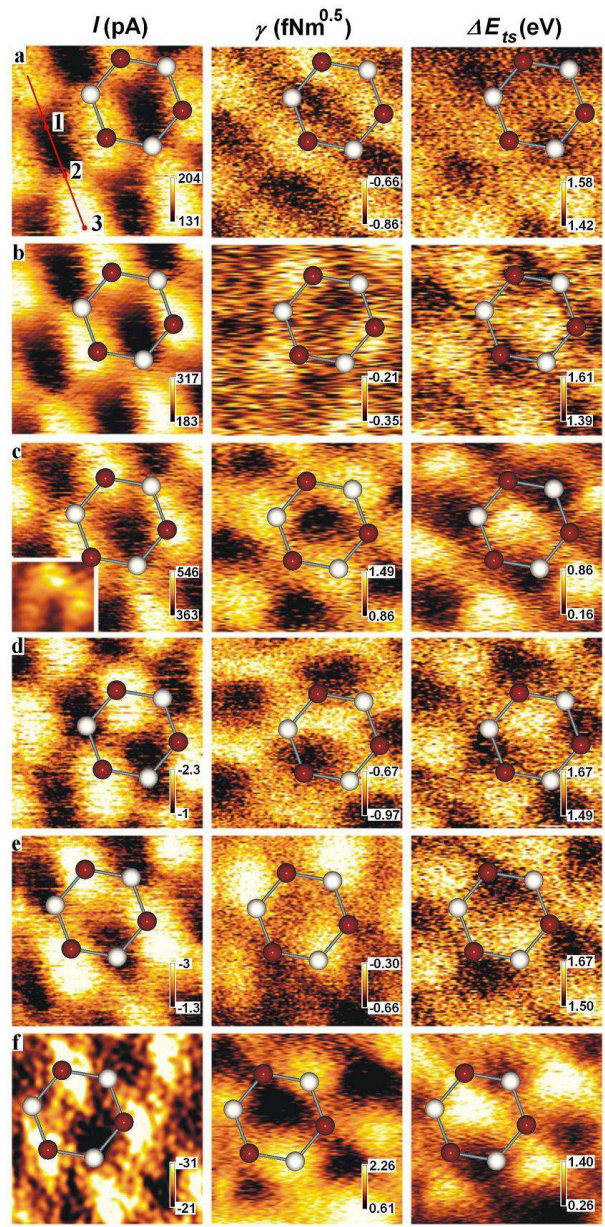}
\caption{ (color) Constant-height measurements of $I$, $\gamma$
and $\Delta E_{ts}$ in attractive and repulsive distance regimes
for a tip bias of 160\,mV (a-c) and -60\,mV (d-f). The hexagons
show the proposed positions of $\alpha$ (white) and $\beta$ (red)
atoms. The brightness is proportional to $|I|$, $\gamma$ and
$\Delta E_{ts}$. The inset in c) left shows a higher harmonic
image \cite{Hembacher:2004}, indicating that the tip state is not
perfectly symmetric with respect to the $z$-axis.} \label{fig4}
\end{figure}
The local spectra at high-symmetry sites were supplemented by
constant height scans at various $z$-positions shown in Fig. 4.
Figures 4a)-c) show $I$, $\gamma$ and $\Delta E_{ts}$ at a tip
bias of 160\,mV in the fully attractive mode at $z\approx
-100$\,pm (a), in a weakly repulsive regime at $z\approx -350$\,pm
(b) and a fully repulsive mode at $z\approx -750$\,pm (c). If the
attractive interaction between tip and sample was only mediated by
the electronic states that contribute to the tunneling current,
the $I$- and $\gamma$-images in Fig. 4a) should be exactly
inverse. Evidently, this is not the case. We therefore conclude,
that electronic states that do not contribute to the tunneling
current may contribute to attractive interaction. In the repulsive
regime shown in Fig. 4c), the repulsion is strongest above the
$\alpha$ sites, almost as strong on the $\beta$ sites, and weakest
on hollow sites, in agreement with with Fig. 3a) and the Pauli
exclusion law. The current has a local maximum on top of the
$\beta$ sites, and $\Delta E_{ts}$ has a pronounced maximum at the
hollow sites. Because of the long time scales involved in damping
measurements, atomically resolved energy loss measurements are
prone to lateral shifts \cite{Gauthier:2002} for fast scanning.
The scanning speeds used in the data of Fig. 4 were $\approx
0.3$\,nm/s, therefore time delays in the acquisition channels are
negligible and the $I$, $\gamma$ and $\Delta E_{ts}$ images match
precisely in forward and backward scans. In our previous
simultaneous STM/AFM measurement on graphite
\cite{Hembacher:2003}, we found a lateral shift of 68\,pm of the
current maxima with respect to the corresponding $\gamma$ maxima.
While the tip is not perfectly symmetric with respect to the
$z$-axis (see caption Fig. 4), it is evidently more symmetric than
in \cite{Hembacher:2003}. Arai and Tomitori \cite{Arai:2004} have
recently shown that force interactions are also a function of
bias. Figure 4d)-f) shows constant-height images at a different
tip bias of -60\,mV. In the attractive regime shown in Fig. 4d),
the $I$- and $\gamma$-images are approximately inverse, i.e. the
local minimum in $\gamma$ coincides with the local maxima in $I$.
In the repulsive regime shown in Fig. 4f), again the repulsion is
strongest above the $\alpha$ sites, almost as strong on the
$\beta$ sites and weakest on hollow sites.

While the distance dependencies of current, force and dissipation
as revealed by constant-height images and local spectroscopic
measurements are qualitatively similar, the contrast observed in
the constant-height images is larger than expected from the local
spectra. The reason for these subtle discrepancies is revealed by
the constant height data shown in Fig. 4. Frequency shift and
current images are different in all distance regimes, and slight
shifts between the high- symmetry points in $I$ and $\gamma$
images are present.

In summary, the spectroscopy experiments show that the
$z$-dependence of force and current is roughly the same as
predicted in Hofer and Fisher's theory \cite{Hofer:2003} for
graphite. However, the constant-height experiments proof that
attractive forces and currents are \emph{not} directly related and
STM and AFM do provide different information. The dissipation
measurements reveal that the theories on atomic friction
introduced by Prandtl \cite{Prandtl:1928} and Tomlinson
\cite{Tomlinson:1929} are the key mechanism for damping on the
atomic scale when imaging soft samples. The spatial resolution
that is possible by scanning probe microscopy scales with the
decay lengths of a physical observables \cite{Stoll:1984}. At the
onset of damping, the decay length of dissipation is only half the
value of the force. This offers an alternate explanation of the
impressive resolution obtained in damping images
\cite{Hug:2002bk}.
\begin{acknowledgments}
This work is supported by the Bundesministerium f\"{u}r Bildung
und Forschung (project EKM13N6918). We thank M. Herz for
discussions and K. Wiedenmann and H. Bielefeldt for help in the
construction of the microscope.
\end{acknowledgments}
\bibliography{2004fjg}
\end{document}